\documentclass{elsart}

\usepackage{natbib}
\usepackage{epsfig}
\usepackage{amssymb}

\begin{document}
\newcommand{\lsun}{\ensuremath{\,{\rm L}_\odot}} 
\newcommand{\ion}[2]{\mbox{#1\,{\scriptsize #2}}}
\newcommand{\mdot}{\mbox{$\dot{M}$}}
\newcommand{\Teff}{\mbox{$T_\mathrm{eff}$}}
\newcommand{\msun}{\ensuremath{\, {\rm M}_\odot}}
\newcommand{\logg}{\mbox{$\log g$}}
\newcommand{\exo}{\mbox{EXO\,0748$-$676}}

\renewcommand{\arraystretch}{1.0}

\begin{frontmatter}

\title{HIRDES -- The High-Resolution Double-Echelle Spectrograph for the World
  Space Observatory Ultraviolet (WSO/UV)}

\author[label1]{K. Werner},
\author[label1]{J. Barnstedt},
\author[label1]{W. Gringel},
\author[label1]{N. Kappelmann},
\author[label2]{H. Becker-Ro{\ss}},
\author[label2]{S. Florek},
\author[label3]{R. Graue},
\author[label3]{D. Kampf},
\author[label3]{A. Reutlinger},
\author[label3]{C. Neumann},
\author[label4]{B. Shustov},
\author[label5]{A. Moisheev},
\author[label5]{E. Skripunov}

\address[label1]{Institut f\"ur Astronomie und Astrophysik, Universit\"at T\"ubingen, Sand 1, 72076 T\"ubingen, Germany}
\address[label2]{Institute for Analytical Sciences, Albert-Einstein-Str. 9, 12484 Berlin,  Germany}
\address[label3]{Kayser-Threde GmbH, Wolfratshauser Str. 48, 81379 M\"unchen, Germany}
\address[label4]{Institute of Astronomy RAS, Pyatnitskaya st. 48, 119017, Moscow, Russia}
\address[label5]{Lavochkin Association, Leningradskoye Shosse, Khimki, 141400, Moscow, Russia}

\begin{abstract}
The World Space Observatory Ultraviolet (WSO/UV) is a multi-national project
grown out of the needs  of the astronomical community to have future access to
the UV range. WSO/UV consists of a single UV  telescope with a primary mirror
of 1.7~m diameter feeding the UV spectrometer  and  UV imagers. The spectrometer
comprises three different spectrographs, two high-resolution echelle
spectrographs (the High-Resolution Double-Echelle Spectrograph, HIRDES) and a
low-dispersion long-slit instrument. Within HIRDES the 102--310~nm spectral band
is split to feed two echelle spectrographs covering the UV range 174--310~nm and
the vacuum-UV range 102--176~nm with high spectral  resolution
($R>50\,000$). The technical concept is based on the heritage of two previous
ORFEUS SPAS missions. The phase-B1 development activities are described in this
paper considering performance aspects, design drivers, related trade-offs
(mechanical concepts, material  selection etc.)  and a critical functional and
environmental test verification approach. The current state of other WSO/UV
scientific instruments (imagers) is also described.
\end{abstract}

\begin{keyword}
Visible and ultraviolet spectrometers \sep Space-based ultraviolet, optical, and
infrared telescopes \sep Astronomical and space-research instrumentation
\PACS 07.60.Rd \sep 95.55.Fw \sep 95.55.-n
\end{keyword}

\end{frontmatter}

\section{Introduction}
The World Space Observatory (WSO/UV) will provide future access to
high-resolution far-UV  spectroscopy. WSO/UV is an international collaboration
led by Russia (Roscosmos) to build a UV (102--310~nm)  mission with capabilities
which are presently and in the near and long-term future unavailable to the
world-wide  astronomical community. The mission is scheduled for
launch in 2010. The planned instrument sensitivity will
exceed that of HST/STIS by a factor of 5--10  and all observing time will be
available for UV astronomy. The present mission design comprises a 1.7~m
telescope. The focal-plane (FP) instruments consist of two high-resolution
spectrographs ($R\sim 55\,000$) covering  the 102--310~nm range and a long-slit
(1''~$\times$~75'') low-resolution ($R\sim 500-5000$) spectrograph. Although
the primary science of the WSO/UV mission is spectroscopy, high
spatial-resolution UV imaging instruments are foreseen. Additionally, a direct
imager which samples the best diffraction-limited resolution of the optical
system is implemented. WSO/UV will be operated like a ground-based telescope,
i.e., it will perform  ``real-time'' operations in an orbit with reduced
visibility constraints (high-Earth orbit).  Overviews of the WSO/UV mission and
its science case were given by \citet{Ba03} and \citet{Go06}, respectively.

\section{The WSO/UV telescope}
\begin{figure}
\begin{center}
\includegraphics[width=0.94\textwidth]{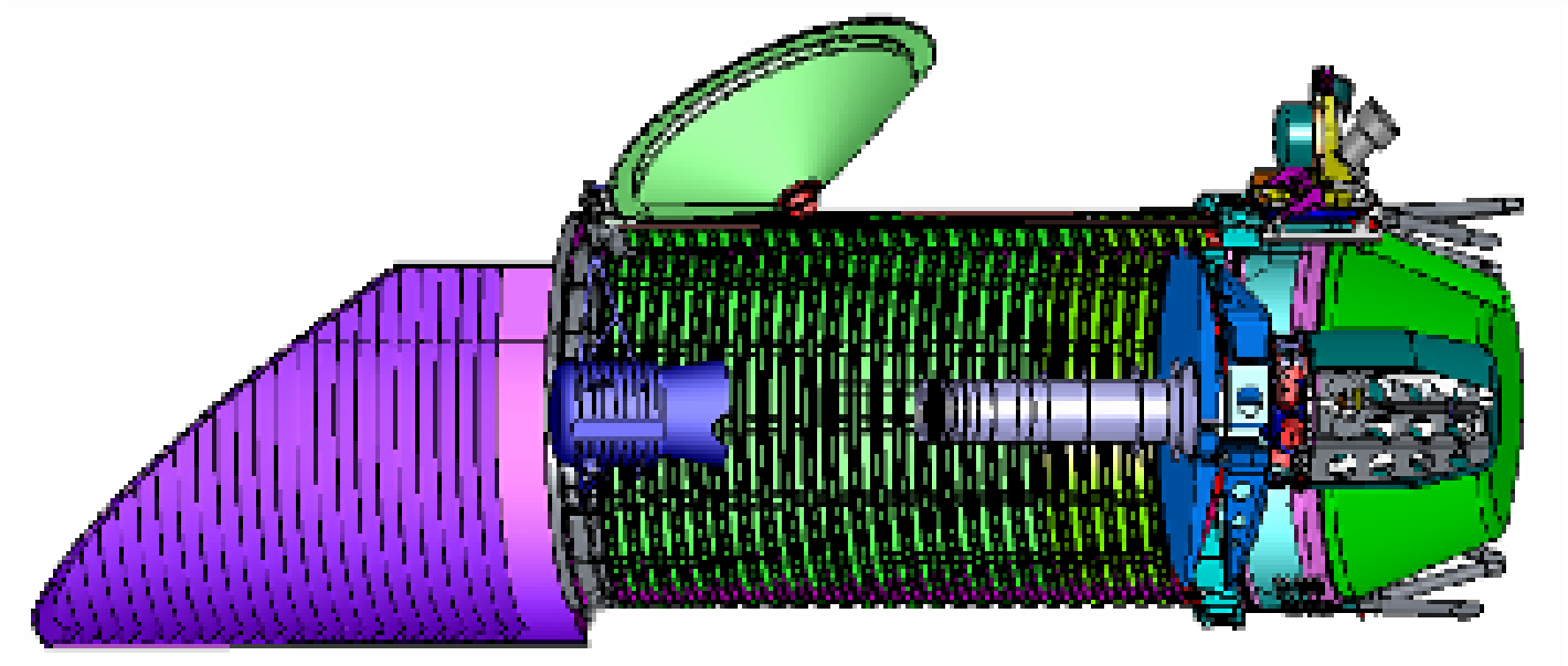}
\includegraphics[width=0.89\textwidth]{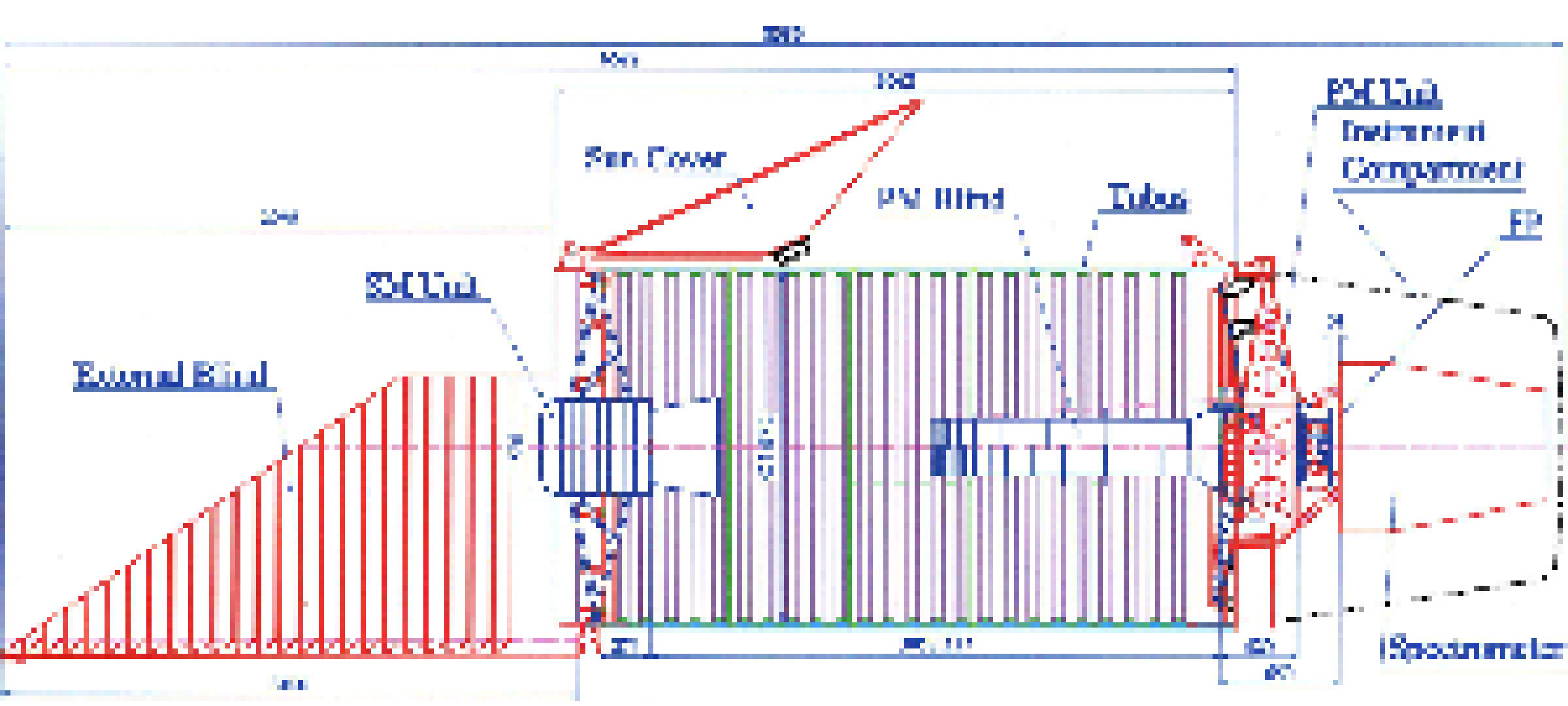}
\caption{The T-170M telescope of WSO/UV.}\label{wernerfig1}
\end{center}
\end{figure}

The heritage for the WSO telescope design is the Russian-led international space
observatory Spectrum-UV  -- a Russian, Ukrainean, German and Italian project --
that was canceled due to funding problems. The WSO/UV  telescope (T-170M) is a
new version of the T-170 telescope designed by Lavochkin Association, Moscow.
Modifications have been made to reduce the weight of the telescope below
1600~kg. The T-170M telescope and its structure are shown in Fig.~\ref{wernerfig1}
with the principal structural elements: the primary and secondary mirrors, and
the instrument compartment. The primary mirror unit (PMU) is the telescope's
main structural element. There are three attachment points of the telescope to
the spacecraft's (S/C) service module (S/M).  The optical bench with the scientific
instrumentation devices and the primary mirror's baffles are mounted on the PMU
frame.  The optical design is a Ritchey-Chr{\'e}tien type with a 1.7~m
hyperbolic mirror. The mirror has an equivalent  focal length of 17.0~m, and a field
of view (FOV) of 30' (\O=150 mm). The optical quality of the main and secondary
mirrors is $\lambda$/30~rms at 633~nm and the angular resolution at the focal
plane is 12.05''/mm. The characteristics of the T-170M telescope are given in
Tab.~\ref{tab1}. The platform for WSO/UV is the same as that developed by
Roscosmos for Spectrum-X-Gamma, the NAVIGATOR  bus, but it will be tailored to
the WSO/UV requirements (Tab.~\ref{tab2}). A Russian Zenith\,2 is the
likely launch vehicle.

\begin{table}
\begin{center}
\caption{Characteristics of the T-170M telescope.}\label{tab1}
\begin{tabular}{ll} 
\noalign{\smallskip} \hline 
Optical system	& Ritchey-Chr{\'e}tien aplanat \\
Aperture diameter	& 1700 mm\\
Telescope f-number	& 10.0\\ 
FOV	& 30' (\O=150 mm)\\
Wavelength range	& 100--310 nm (+visible)\\
Primary wavelength	& 200 nm\\
Optical quality	& Diffraction optics at the FOV center\\
Mass	& 1570 kg (1600 kg with adapter truss)\\         
Size	& 5.67\,m\,$\times$\,2.30\,m (transport); 8.43\,m\,$\times$\,2.3\,m (operational)\\
\hline 
\end{tabular}
\normalsize
\end{center}
\end{table}

\begin{table}
\begin{center}
\caption{Characteristics of the NAVIGATOR bus.}\label{tab2}
\begin{tabular}{ll}
\noalign{\smallskip} \hline 
Platform (S/M) &	Navigator \\ 
Mass of S/C with propellant &	2900 kg  \\ 
S/M mass		 & 1300 kg \\         
S/M propellant mass	 & 150 kg \\         
S/M propellant	 & Hydrazine \\         
Payload (P/L) mass & 	1600 kg \\         
Pointing accuracy with star sensors	 & 4 (2)' \\
Accuracy of pointing and stabilization with the FGS	 & 0.1''\\
Slewing rate	 & Up to 0.1$^\circ$/sec \\
Maximum exposure time	 & 30 hours \\ 
Download of scientific data	 & Up to 1 Mbit/sec \\
H/K data transmission rate	 & Up to 32 Kbit/sec \\
Electric power (EP) available for P/L	 & 750 W \\
EP for science instrument for the FP compartment	 & 300 W \\
Voltage of electric power supply	 & 27$\pm$1.35 V \\ \hline 
\end{tabular}
\normalsize
\end{center}
\end{table}

\begin{figure}
\begin{center}
\includegraphics[width=0.79\textwidth]{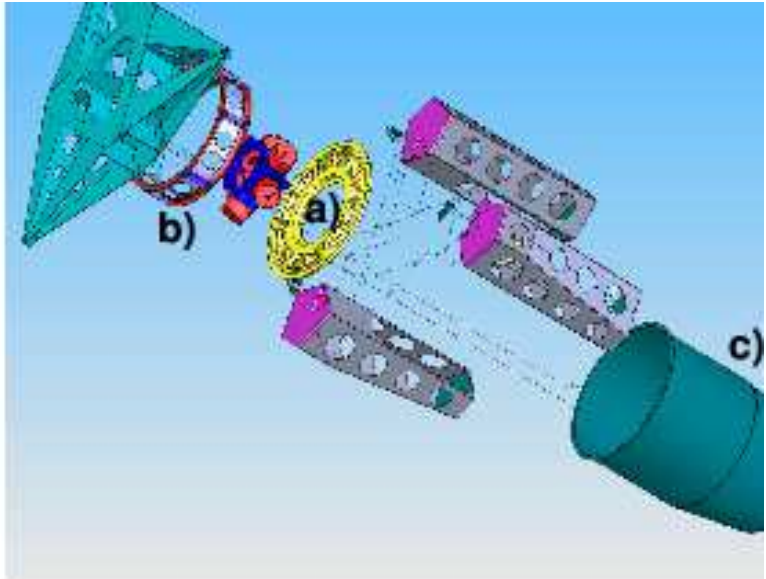}
\caption{The WSO/UV spacecraft interfaces, i.e., a) optical bench plate,
  b) external
instrument plate, and c) protective case.}\label{wernerfig2}
\end{center}
\end{figure}

\section{The high-resolution double-echelle spectrograph}

The UV spectrometer comprises three different single spectrographs, two
high-resolution echelle spectrographs  -- the High-Resolution Double-Echelle
Spectrograph (HIRDES) -- and a low-dispersion long-slit instrument. The  HIRDES
spectral band (102--310~nm) is split and fed into two echelle spectrographs
covering the UV range 174--310~nm (UVES channel) and vacuum-UV range 102--176~nm
(VUVES channel) with a high spectral  resolution of $R>\,50\,000$. Each
spectrograph encompasses a standalone optical bench structure with a fully
redundant  high-speed MCP detector system, the optomechanics and a network of
mechanisms with different functionalities.  After a HIRDES assessment study for
the Spectrum-UV mission a phase-A study was performed by Jena-Optronik, Jena, in
2001. The main goal of this study was the rearrangement of the echelle
spectrographs and the long-slit spectrograph due to the fact that HIRDES was
considered the only focal-plane spectrograph in the instrumental bay of
WSO/UV. HIRDES is located in the S/C instrument compartment with mounting
interfaces  to the main S/C structure (optical bench) and it is covered by a
protective case (Fig.~\ref{wernerfig2}). The associated electronic boxes are
located separately at the protective case and the electronic-box panels are
provided by the S/C.

\section{Phase-B1 study of the spectrographs}

\begin{figure}
\begin{center}
\includegraphics[width=0.49\textwidth]{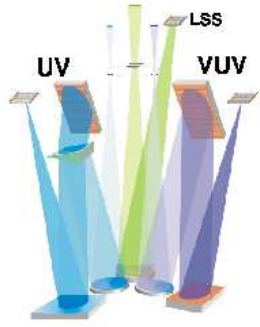}
\caption{Optical layout for the UV spectrograph with an echelle grating
  (40~lines/mm, 70$^\circ$ blaze angle) and prism (fused silica, 12$^\circ$);
  and optical layout for the 
VUV spectrograph: dispersion by echelle grating 
(65~lines/mm, 71$^\circ$) and order separation by cross disperser 
(on-focus mirror, 625~lines/mm). A preliminary optical layout of 
the long-slit spectrograph (LSS) is added.}\label{wernerfig3}
\end{center}
\end{figure}

A phase-B1 study was performed by Kayser-Threde GmbH, Munich, in close
cooperation with Lavochkin Association, Moscow. It was started in 2005
and finished in April 2006. The Institute of Analytical  Sciences in Berlin was
responsible for the optical layout of the spectrographs. The optical layout of
the long-slit spectrograph (LSS) was not part of this study. A phase-A study for
the LSS is currently being performed in China.  The main spectrometer requirements are
given in Tab.~\ref{tab3} and the optical design of HIRDES is shown in
Fig.~\ref{wernerfig3}.

\begin{table}
\begin{center}
\caption{General requirements for UV and VUV echelle spectrographs (UVES and VUVES).}\label{tab3}
\begin{tabular}{ll} 
\noalign{\smallskip} \hline 
Parameter   & 	Baseline requirements \\\hline 
Wavelength coverage VUVES& 102--176 nm\\
Wavelength coverage UVES& 174--310 nm \\
Spectral resolution	 & $R>50\,000$\\
Minimum sensitivity SNR=10  in 10 h & m$_{\rm VUV}=16$; m$_{\rm UV}=18$  \\
Minimum sensitivity SNR=100 in 10 h & m$_{\rm VUV}=11$; m$_{\rm UV}=13$  \\
Limit loads in all axes w/o SF & 	15 g (tbc)\\
Stiffness (1st fundamental eigenfrequency)	 & $\tt>$ 40 Hz\\
Envelope 	 & Protective case\\
Mass	 & 155 kg\\
Power & 	150 W\\
Data rate (downlink)	 & 1.6 Mbit/sec\\\hline 
\end{tabular}
\end{center}
\end{table}

\begin{figure}
\begin{center}
\includegraphics[width=0.49\textwidth]{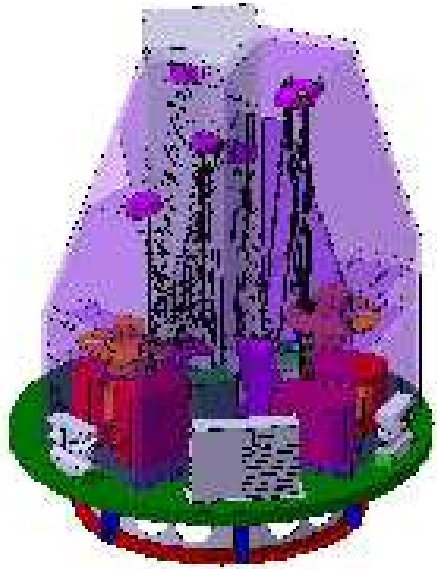}
\caption{Schematic overview of the UV and VUV spectrometers and electronic boxes.}\label{wernerfig4}
\end{center}
\end{figure}

The UV and VUV  spectrographs are equipped with an in-field fine guidance system
(IFGS). The IFGS is used for a multi-step pointing system to stabilize the spectral image
of the spectrometers. The IFGS sensors are placed near the entrance
slits. For the focal-plane array detectors of the IFGS a
selection between two detector concepts (CCD, CMOS) was performed. CMOS
detectors were selected because they are less environment sensitive (radiation,
EMC) and  easier to operate.
The design envelope of the spectrometers and external electronic boxes is mainly
determined by the S/C interfaces  (protective case and optical bench). An
overview of the HIRDES is given in Fig.~\ref{wernerfig4}.

\begin{figure}
\begin{minipage}[b]{0.5\linewidth} 
\centering
\includegraphics[width=6cm]{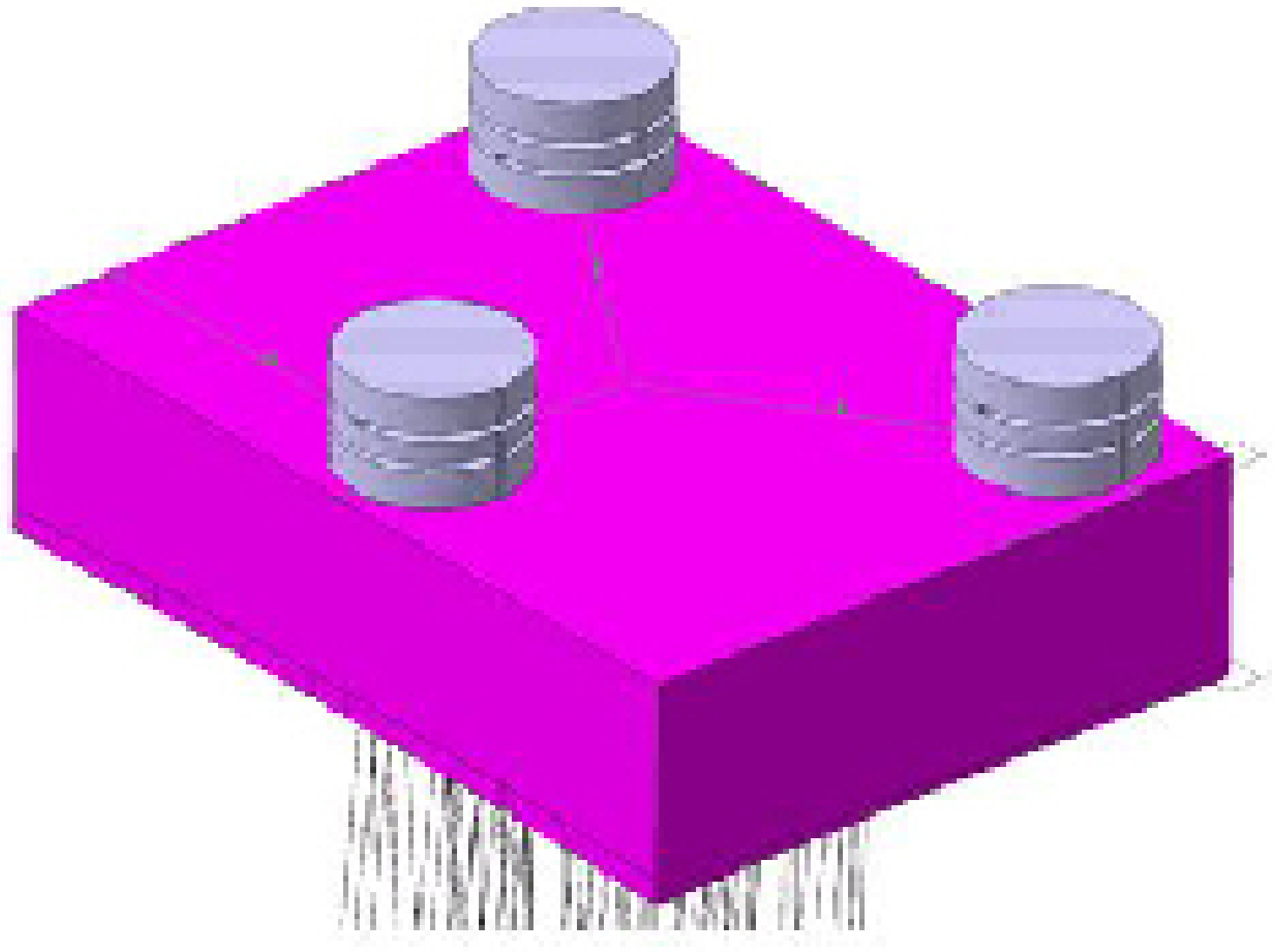}
\caption{Isostatic mirror suspension.}\label{wernerfig5}
\end{minipage}
\hspace{0.1cm} 
\begin{minipage}[b]{0.5\linewidth}
\centering
\includegraphics[width=6cm]{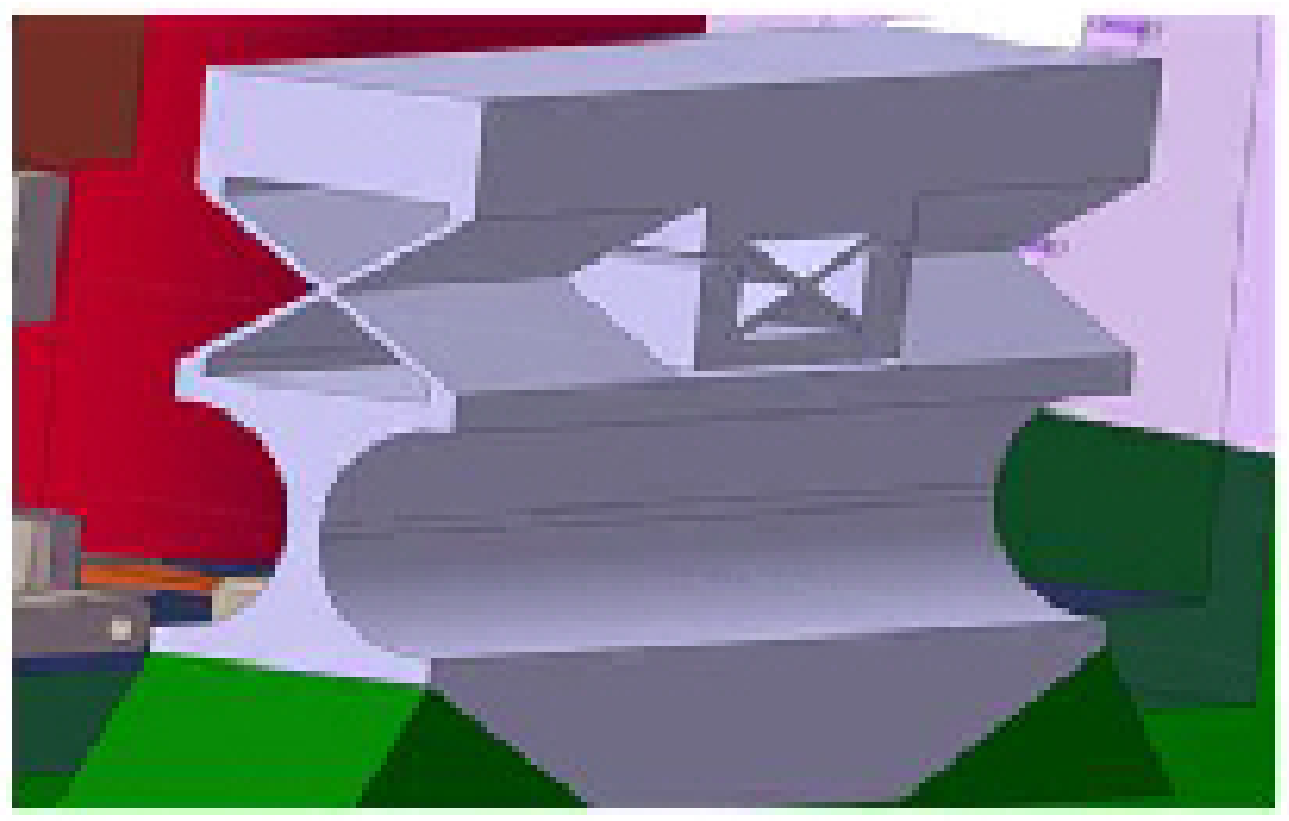}
\caption{Invar suspension with stainless steel flexural blades.}\label{wernerfig6}
\end{minipage}
\end{figure}

The flexural mounts are made from Invar and shall be bonded into the mirror
substrates as given in Fig.~\ref{wernerfig5}. This technique  was successfully
qualified for the ORFEUS telescope. The HIRDES suspension is configured to
establish  the isostatic mounting of the three spectrometers at the S/C optical
bench. Considering the limited envelope and  the high mechanical loads Invar
pads with stainless steel flexurable links with high strength offers the
best  performance in terms of stiffness, strength and minimized thermal
conductivity (ORFEUS heritage), see Fig.~\ref{wernerfig6}.

\begin{figure}
\begin{minipage}[b]{0.5\linewidth} 
\centering
\includegraphics[width=6cm]{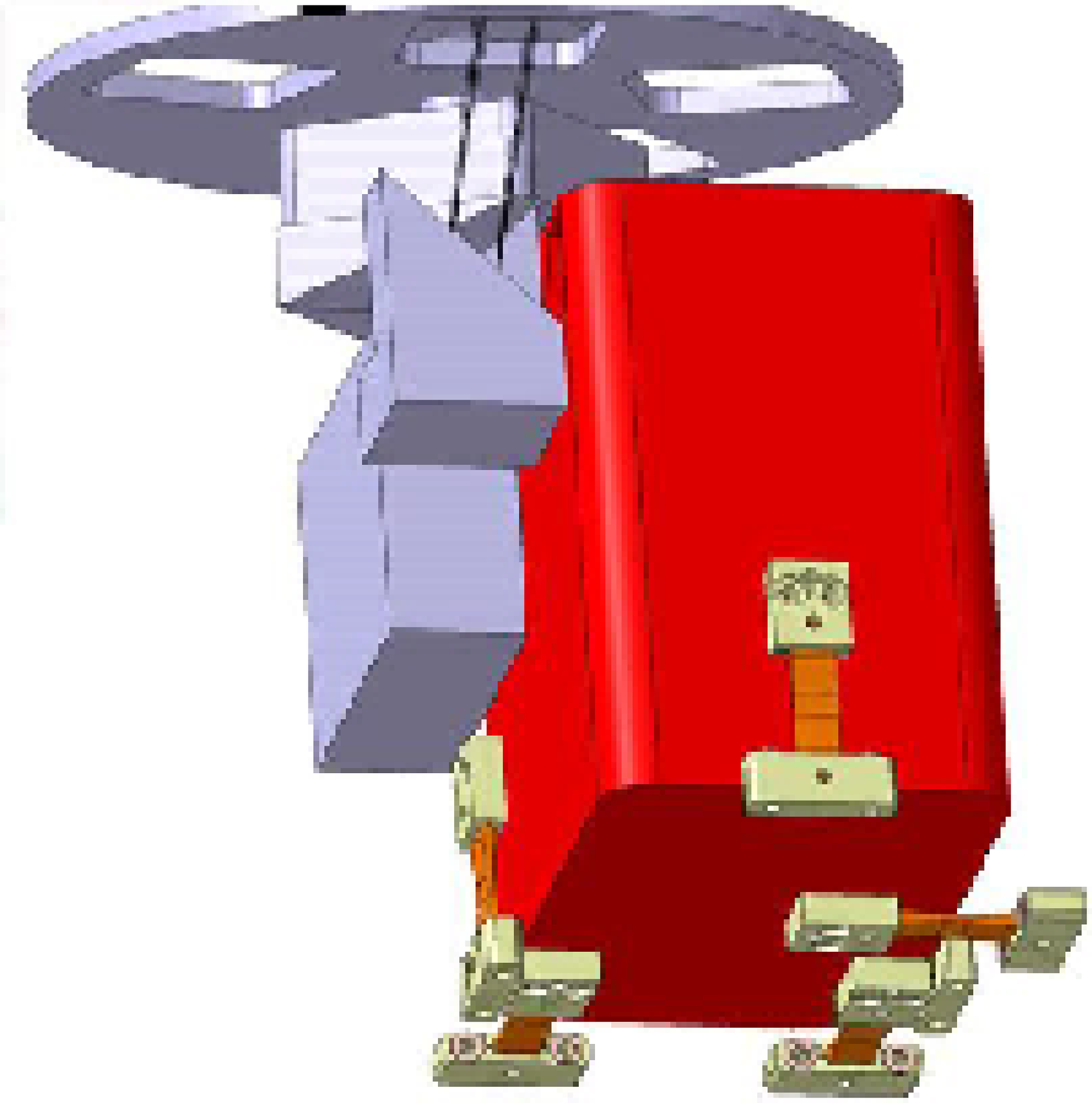}
\caption{MCP detector box with isostatic suspension made from carbon fibre, and
  with a filter wheel for grey filters.}\label{wernerfig7}
\end{minipage}
\hspace{0.1cm} 
\begin{minipage}[b]{0.5\linewidth}
\centering
\includegraphics[width=6cm]{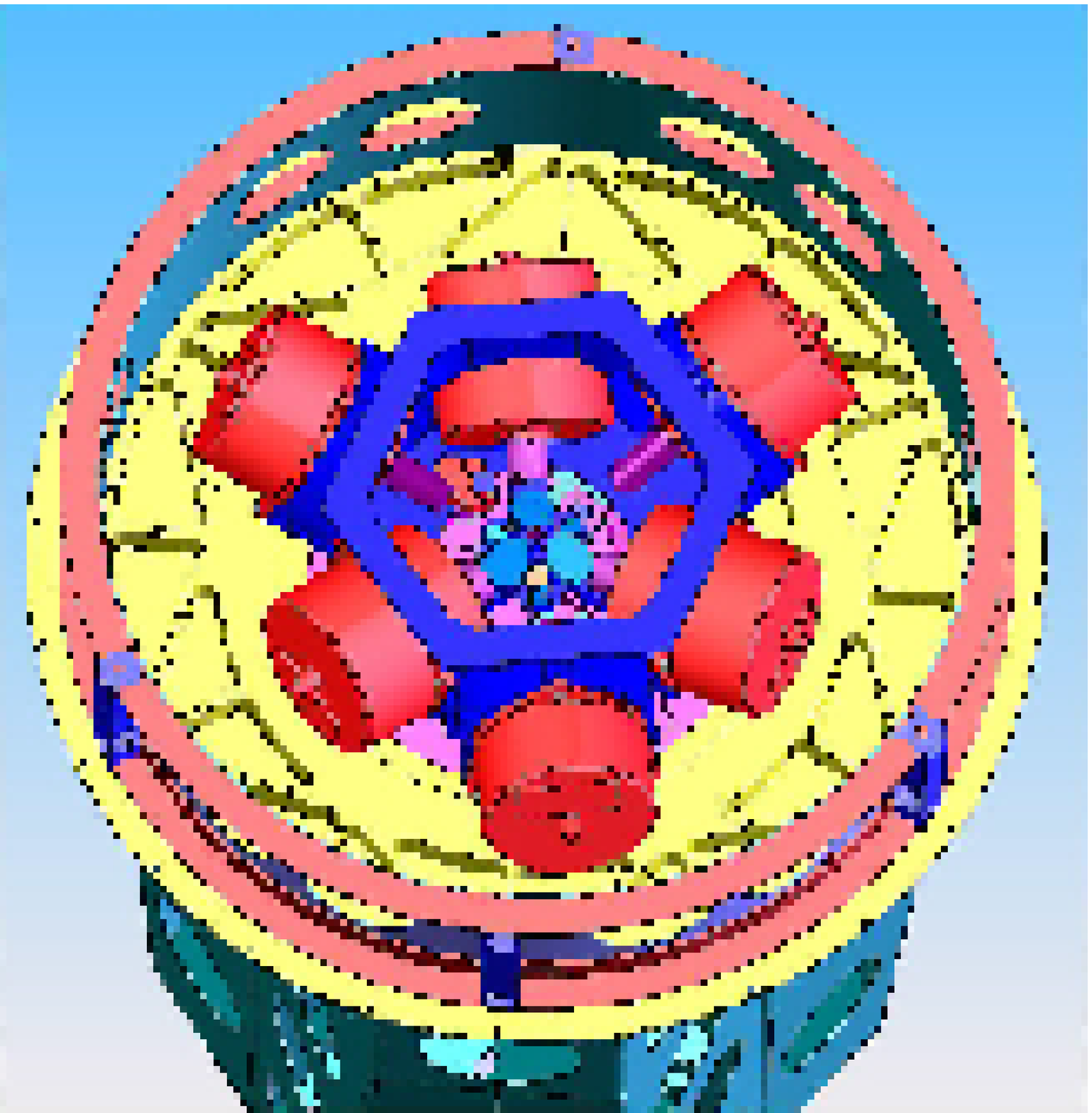}
\caption{Preliminary arrangement of the focal camera unit.}\label{wernerfig8}
\end{minipage}
\end{figure}

The following mechanisms are foreseen for both UVES and VUVES: vacuum shutter
mechanism for the detectors, servo-mirror  mechanism to switch between normal operating and
redundant detector heads, and a mechanism for grey filters
(Fig.~\ref{wernerfig7}) which will be used to observe bright targets.

The HIRDES and IFGS detector suspensions are configured to establish the isostatic
mounting of the boxes and additionally a thermal decoupling of these boxes from
the spectrometer structure. By separating the directions a strut solution was
found, where every strut supports one degree of freedom and is flexible for the
others (Fig.~\ref{wernerfig7}). The struts are made of CFRP (Herschel-PACS
heritage).

The material of mirrors to be used must be very homogeneous. Several different
types of materials were traded with respect to environmental and performance
requirements.  Fused silica mirrors with Invar flexible mounts (ORFEUS heritage)
were selected due to superior performance in terms of light-weighting, bonding
techniques, surface quality and manufacturing costs.

Very low mechanical tolerances are required due to high optical stability
requirements, namely translation: 10--50 $\mu$m, and rotation: 2--5''.  The
trade-off of the following candidate materials has given the technical and
related costs features: composite materials, aluminum and ceramics.  The
candidate materials were investigated. Aluminum seems to be very critical with
respect to the thermal stability requirements. Due to quasi-ambient
temperatures, high accuracy, cost efficiency and specific experience the
selected baseline for the primary structure of the spectrometers was the concept
based on CeSiC. The thermostable CeSiC structure is not sensitive to thermal
gradients. No mechanisms with high functional, technical and operational
complexity and dedicated system costs are necessary. Therefore the required
spectral resolution is achieved without active control of optical elements, and
the complex collimator and echelle-grating mechanisms  for focusing (introduced
in the assessment and phase-A studies) are no longer necessary.

The architecture of the structural design of UVES and VUVES spectrometer
structures provides the following features: standalone structure, independent
assembly and alignment S/C accommodation, independent functional performance
testing.  The envelope of the UVES and VUVES spectrometers covers 240$^\circ$,
the envelope of the LSS covers 120$^\circ$ of the instrument compartment around
the optical axis.  The following features can be
highlighted with respect to the structural design:

\begin{itemize}
\item	Monolithic structure made from CeSiC
\item	Cost-effective light-weight approach; fabrication technology available
\item	Reduced manufacturing tolerances (procurement costs) due to standardized adjustment devices for each optical component
\item	Isostatic suspension for the HIRDES spectrometers to the S/C interface
\item	Isostatic suspensions, i.e., thermo-mechanical decoupling for the detectors from the main spectrometer housings
\end{itemize}

\section{The focal plane imagers}

Although the primary aim of WSO/UV is spectroscopy, there is an
important role for high-resolution imaging. The focal camera unit
(FCU) will include:

\begin{itemize}
\item	An optical camera (OC), working at the best diffraction-limited
resolution with the largest FOV which is possible to
accommodate; this camera is intended to perform astrometry in crowded fields.
\item	Two UV imagers: one f/50 camera with a resolution of $\sim 0.03$''/px and
\mbox{$\sim 1.2$'} FOV (long focus, LF), and one f/10 camera with $\sim 0.15$''/px resolution
and $\sim 6$' FOV (short focus, SF). Each of these cameras is equipped with one or
two filter wheels in order to accommodate passband filters.
\end{itemize}

The possibility to accommodate redundant UV and optical cameras in the FCU has
to be evaluated during the phase-A/B1 study (Fig.~\ref{wernerfig8} shows a possible
design and Tab.~\ref{tab4} summarizes the specifications), which shall be completed in
Italy in 2007.  The final choice of detectors -- MCP and/or CCD -- will be a
task of this study. However, from a preliminary assessment study the baseline
choice is MCP for the UV cameras and CCD for the optical camera.

\begin{table}
\begin{center}
\caption{Specifications for the focal cameras.}\label{tab4}
\begin{tabular}{ccccrr} 
\noalign{\smallskip} \hline 
{Camera} & {Range} & {Focal ratio} & FOV &  {PSF sampling}   &  {Resolution} \\\hline 
SF	 &     UV	 & f/10	 & 6.0'	        & 0.15''/px	 & 0.3''\\
LF	 &     UV	 & f/50	 & 1.2'	        & 0.03''/px	 & 0.1''\\
OC	 &    visible	 & tbd	 & tbd    &  $\le$0.03''/px	 & $\le$0.1''\\\hline 
\end{tabular}
\end{center}
\end{table}

\section{Conclusion}

The WSO/UV mission will provide high-resolution UV spectroscopy and imaging. We have
presented parts of the phase-B1 study of the main instrument of the WSO/UV
mission, the High-Resolution Double-Echelle Spectrograph HIRDES. With the choice
of a ceramic material (CeSiC) for the structure of the optical benches of the
individual standalone spectrographs, the high spectral resolution can be
achieved without active control of optical elements. Furthermore it is not
necessary to install complex focusing mechanisms.  The introduced mirror,
detector and spectrometer suspensions are already successfully qualified by
other space missions. 

{\bf Acknowledgement} The phase-B1 study of HIRDES was supported and financed
by the Bundesministerium f\"ur Wirtschaft und  Technologie, Deutsches Zentrum
f\"ur Luft und Raumfahrt e.V. (DLR), through grant FKZ 50\,QV\,0503.

\end{document}